# Far-field optical microscope with nanometer-scale resolution

Igor I. Smolyaninov, Christopher C. Davis, Jill Elliott* & Anatoly V. Zayats*

*Department of Electrical and Computer Engineering, University of Maryland, College Park, MD 20742, USA*

**School of Mathematics and Physics, Queen's University of Belfast, Belfast BT7 1NN, United Kingdom*

**The resolution of far-field optical microscopes, which rely on propagating optical modes, is widely believed to be limited because of diffraction to a value on the order of a half-wavelength $\lambda/2$ of the light used. Although immersion microscopes have slightly improved resolution on the order of $\lambda/2n$, the increased resolution is limited by the small range of refractive indices $n$ of available transparent materials. Here we demonstrate a new far-field optical microscope design, which is capable of reaching nanometer-scale resolution. This microscope uses the fact that the effective refractive index $n_{eff}$ of a planar dielectric lens or mirror placed on a metal surface may reach extremely large values, up to $10^3$, as seen by propagating surface optical modes (plasmons). In our design a magnified planar image produced originally by surface plasmons in the metal plane is viewed by a regular microscope. Thus, the theoretical diffraction limit on resolution is pushed down to nanometer-scale $\lambda/2n_{eff}$ values. Used in reverse, such a microscope may become an optical lithography tool with nanometer-scale spatial resolution.**

Optical microscopy is one of the oldest research tools. It dates back to 1609 when Galileo Galilei developed an *occhiolino* or compound microscope with a convex and a concave lens. Although various electron and scanning probe microscopes have long surpassed it in resolving power, optical microscopy remains invaluable in many fields of science. The reason for the limited resolution of an optical microscope is diffraction and, ultimately, the uncertainty principle: a wave can not be localized much tighter than half of its vacuum wavelength $\lambda/2$. Immersion microscopes [1] introduced by Abbe in the 19th century have slightly improved resolution on the order of $\lambda/2n$ because of the shorter wavelength of light $\lambda/n$ in a medium with refractive index $n$. However, immersion microscopes are limited by the small range of refractive indices $n$ of available transparent materials. For a while it was believed that the only way to achieve nanometer-scale spatial resolution in an optical microscope is to beat diffraction, and detect evanescent optical waves in very close proximity to a studied sample using a scanning near-field optical microscope [2]. Recently, some alternative approaches based on Pendry's idea of a "perfect lens" made from an artificial negative refractive index material have been introduced [3]. In this theoretical scheme a high resolution image would be obtained by amplified evanescent waves. However, such an image would be observable only in the near-field of a perfect lens, and would require an auxiliary near-field microscope. Although many fascinating results are being obtained with near-field optics, such microscopes are not as versatile and convenient to use as regular far-field optical microscopes. For example, an image of a near-field optical microscope is obtained by point-by-point scanning, which is an indirect and a rather slow process. In this letter we present a new far-field optical microscope design, which is capable of direct visualization of nanometer-scale samples. We believe that the new technique will lead to numerous breakthroughs in biological imaging and subwavelength lithography.

Very recently it was realized [4] that a dielectric droplet on a metal surface which supports propagation of surface optical modes may have an extremely large effective refractive index as seen by these modes (these modes are usually called surface plasmons or surface plasmon-polaritons). The properties of these modes and convenient ways to excite them are described in detail in [5]. The wave vector of surface plasmons is defined by the expression

$$k_p = \frac{\omega}{c}\left(\frac{\varepsilon_d \varepsilon_m}{\varepsilon_d + \varepsilon_m}\right)^{1/2} \quad (1)$$

where $\varepsilon_m(\omega)$ and $\varepsilon_d(\omega)$ are the frequency-dependent dielectric constants of the metal and dielectric, respectively [5]. Under the resonant condition

$$\varepsilon_m(\omega) = -\varepsilon_d(\omega) \quad (2)$$

both phase and group velocities of the surface optical mode tend to zero. This means that the wavelength $\lambda_p$ of such modes becomes very small just below the optical frequency defined by equation (2), or in other words, the effective refractive index of the dielectric $n_{eff}$ becomes extremely large as seen by the propagating surface plasmons in this frequency range. As a result, a small droplet of liquid dielectric on the metal surface becomes a very strong lens for surface plasmons propagating through the droplet from the outside. On the other hand, the droplet boundary becomes an extremely efficient mirror for surface plasmons propagating inside the droplet at almost any angle of incidence due to the total internal reflection (this leads to the "black hole" analogy described in [4]).

Let us consider a far-field two-dimensional (2D) optical microscope made of such droplets or any other appropriately shaped planar dielectric lenses and/or mirrors as shown in Fig.1(a). Since the wavelength of surface plasmons $\lambda_p$ observed in experiments may be as small as a few nanometers [6] (hence $n_{eff}$ may reach extremely large values up to $10^3$), the diffraction limit of resolution of such a 2D microscope may approach $\lambda_p/2$ or $\lambda/2n_{eff}$. Theoretically it may reach a scale of a few nanometers. If a sample under investigation is forced to emit propagating surface plasmons or if it is illuminated by propagating plasmons, these plasmons may produce a 2D magnified image of the sample in the appropriate location on the metal surface. Because of the metal surface roughness and the Raleigh scattering in the dielectric, the propagating plasmons are constantly scattered into normal photons propagating in free space. As a result, the plasmon-produced far-field 2D image on the metal surface may be visualized by a normal optical microscope. The image brightness far exceeds the background of scattered plasmons in other areas of the 2D microscope, and in addition, a fluorescence scheme of surface plasmon field visualization by a far-field optical microscope may be used [7]. Thus, the goal of a 2D microscope design is to have sufficiently high 2D image magnification, so that all the 2D image details would be larger than the $\lambda/2$ resolution limit of the normal optical microscope. As a result, a far-field optical microscope with nanometer-scale resolution will be produced. Below we describe an experimental realization [8] of such an optical superresolution far-field microscope, and report experimental proof of its resolution of at least 60 nm, which is equal to approximately $\lambda/8$, and far supersedes resolution of any other known far-field optical microscope.

In a scheme similar to one described earlier in [4], glycerin microdroplets have been used as 2D optical elements in the design of the microscope. The dielectric constant of glycerin $\varepsilon_g=2.161$ is ideally suited for experiments performed on a gold surface within the wavelength range of the laser lines of an argon-ion laser (Fig.1(b)). At the $\lambda_0=502$ nm line the real part of the gold dielectric constant is $\varepsilon_m=-2.256$ [9].

According to equation (1) the corresponding surface plasmon wavelength inside glycerin is $\lambda_p$=69.8 nm, and the effective refractive index of glycerin is $n_{eff} = \lambda_0/\lambda_p =$ 7.14. On the other hand, surface plasmons are not excited at the 458 nm laser line. As a result, frequency-dependent anomalously high spatial resolution of the microscope around 502 nm laser line may be conveniently demonstrated.

In our experiments the samples were immersed inside glycerin droplets on the gold film surface. The droplets were formed in desired locations by bringing a small probe wetted in glycerin into close proximity to a sample. The size and location of the droplets were determined by the probe movement under the visual control, using a regular microscope. Our droplet deposition procedure allowed us to form droplet shapes, which were reasonably close to parabolic (Figs.2,3). Thus, the droplet boundary was used as a 2D parabolic mirror for propagating surface plasmons excited inside the droplet by external laser illumination (Fig.1(a)). Since the plasmon wavelength is much smaller than the droplet sizes, the image formation in such a mirror can be analyzed by simple geometrical optics in two dimensions as shown in Fig.2. In this experiment a set of artificial pinholes in gold film was produced inside a glycerin droplet by scratching the film with a sharp STM tip, and used as a test sample. Such pinholes are known to emit propagating surface plasmon beams [10]. The edges of the scratch (pinholes *A* and *B* in Fig.2(a)) appear to be brighter than the rest of the sample. The position of the focus of the parabolic droplet edge was found from the known edge geometry, and shown by a green dot in Fig.2(b). Given the found focus location, ray optics was used in Fig.2(b) to show that the brighter edges of the scratch are imaged into dimmer points *a* and *b*, which are visible near the parabola focus location. As expected, the distance *ab* is smaller than the distance *AB*. Thus, imaging properties and image demagnification by a glycerine lens has been shown in this experiment. Such reduced surface plasmon images may be used in subwavelength optical lithography.

Periodic nanohole arrays first studied by Ebbesen *et al.* [11] also appear to be ideal test samples for this microscope. Illuminated by laser light such arrays produce propagating surface waves, which explains the anomalous transmission of such arrays at optical frequencies. Fig.3 shows various degrees of 2D image magnification obtained with a 30x30 µm$^2$ rectangular nanohole array with 500 nm hole spacing described in [12] and used as a test sample. In general, smaller glycerine droplets produced higher magnification in the images. Approximate reconstructions of the images via ray tracing are shown next to each experimental image. Individual nanoholes of the array are shown as individual dots in theoretical images. Comparison of Fig.3(c) and Fig.3(d) indicates that the rows of nanoholes separated by 0.5 µm are resolved in the image (c) obtained using only a 10x magnification of the conventional microscope, while comparison of Fig.3(e) and Fig.3(f) indicates that individual 150 nm diameter nanoholes separated by 0.5 µm gaps are resolved in the image (e) obtained at 502 nm. These individual nanoholes are located in close proximity to the focus of the droplet/mirror, and hence experience the highest image magnification. The cross section (f) through the row of nanoholes in (e) indicates edge resolution of at least 100 nm. Images taken at different laser wavelengths in the geometry shown in Fig.3(e) indicate drastic reduction in resolution away from the 502 nm wavelength (compare Fig.3g and 3h).

Another resolution test of the microscope has been performed using a 30x30 µm$^2$ array of triplet nanoholes (100 nm hole diameter with 40 nm distance between the hole edges) shown in Fig.4(c). This array was imaged using a glycerine droplet shown in Fig.4(a). The image of the triplet array obtained at 515 nm is shown in Fig.4(b)

(compare it with an image in (f) calculated using geometrical optics). Although the expected resolution of the microscope at 515 nm is somewhat lower than at 502 nm, the 515 nm laser line is brighter, which allowed us to obtain more contrast in the 2D image. The least-distorted part of the image shown at higher zoom of the conventional optical microscope in Fig.4(d) clearly visualizes the triplet nanohole structure of the sample. The analysis of the cross section through the line of double holes in Fig.4(e) clearly indicate spatial resolution of at least 60 nm. Thus, at least 60 nm ($\lambda/8$) spatial resolution of the microscope is clearly demonstrated. Theoretical resolution of such microscope may reach the nanometer scale. Such a microscope has the potential to become an invaluable tool in medical and biological imaging, where far-field optical imaging of individual viruses and DNA molecules may become a reality. It allows very simple, fast, robust and straightforward image acquisition. Moreover, water droplets on a metal surface can be used as elements of 2D optics in measurements where aqueous environment is essential for biological studies. In addition, used in reverse such a microscope may be used in nanometer-scale optical lithography. Both developments would potentially revolutionize their respective fields.

Figure Captions

Fig.1 (a) Surface Plasmon Immersion Microscope: Surface plasmons are excited by laser light and propagate inside a parabolic-shaped droplet. Placing a sample near the focus of a parabola produces a magnified image in the metal plane, which is viewed from the top by a regular microscope. Used in reverse, this configuration may be used in subwavelength optical lithography. (b) Sketch of the Ar-ion laser lines positions with respect to the dispersion curve of plasmons on the gold-glycerine interface. At 502 nm glycerine has a very large effective refractive index for surface plasmons. Also shown are the approximate locations of other guided optical modes inside the thin layer of glycerine.

Fig.2 Image demagnification by a glycerine lens: Ray optics is used to show that the brighter edges (points *A* and *B* in (a)) of an artificial scratch inside a glycerine droplet are imaged into points *a* and *b*, which are located near the geometrically defined position of the focus (shown by the green dot in (b)) of the parabolic mirror formed by the left edge of the droplet.

Fig.3 2D images of a 30x30 $\mu m^2$ rectangular nanohole array with 500 nm hole spacing, which are formed in various droplets. Approximate reconstructions of the images via ray tracing are shown next to each experimental image. Individual nanoholes of the array are shown as individual dots in the theoretical images. Comparison of (e) and (f) indicates that individual nanoholes are resolved in the image (e) obtained at 502 nm. The cross section (g) through the row of nanoholes in (e) indicates edge resolution of at least 100 nm obtained at 502 nm. The spatial resolution is lost in measurements at 458 nm (h) in which surface plasmons are not excited.

Fig.4 Resolution test of the microscope. The array of triplet nanoholes (c) is imaged using a glycerine droplet shown in (a). The image of the triplet array obtained at 515 nm is shown in (b). The least-distorted part of the image shown at higher zoom in (d), the cross section through the line of double holes (e), and the comparison of image (b) with theoretically calculated image (f) clearly prove the resolving of the triplet structure.

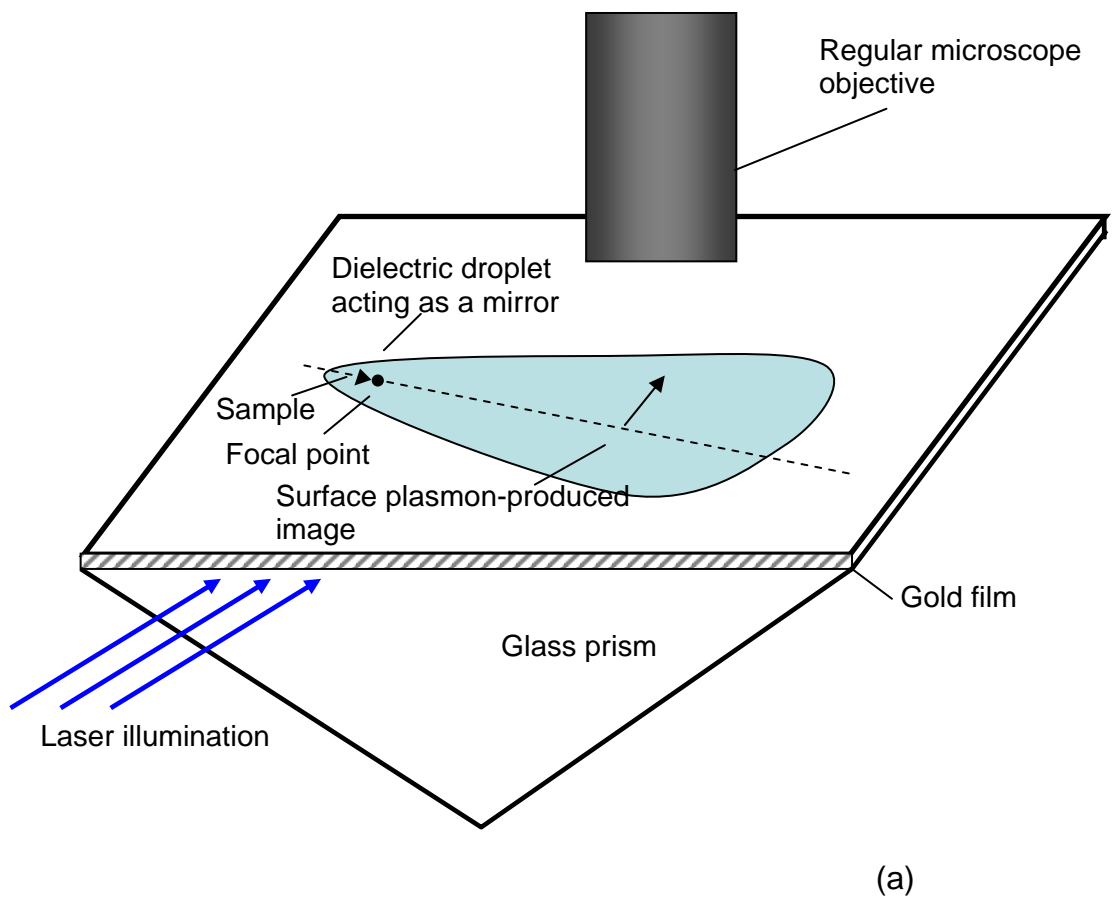

(a)

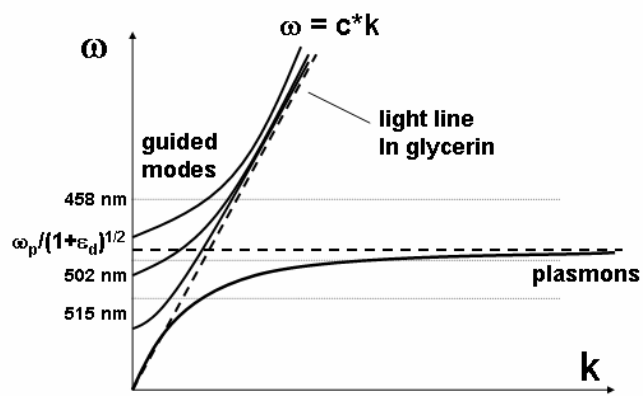

(b)

Fig.1

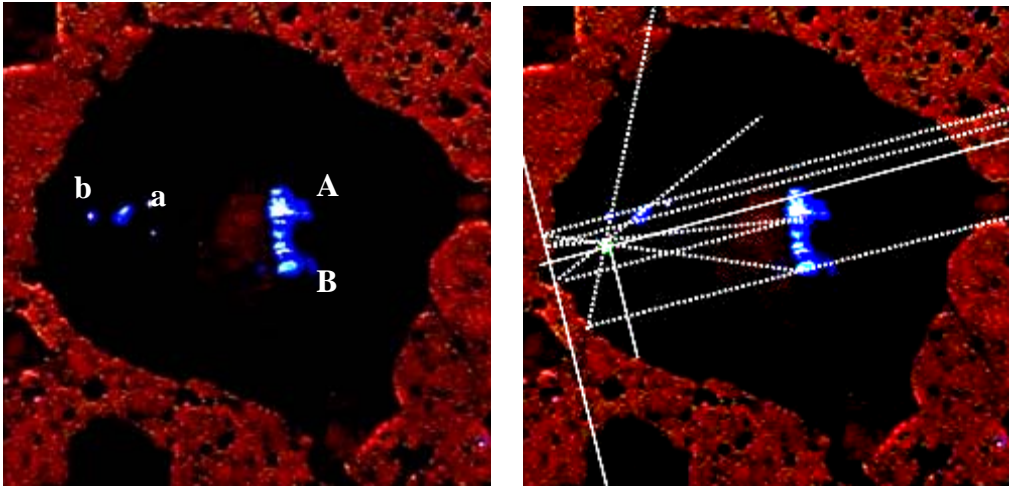

(a)  (b)

Fig. 2

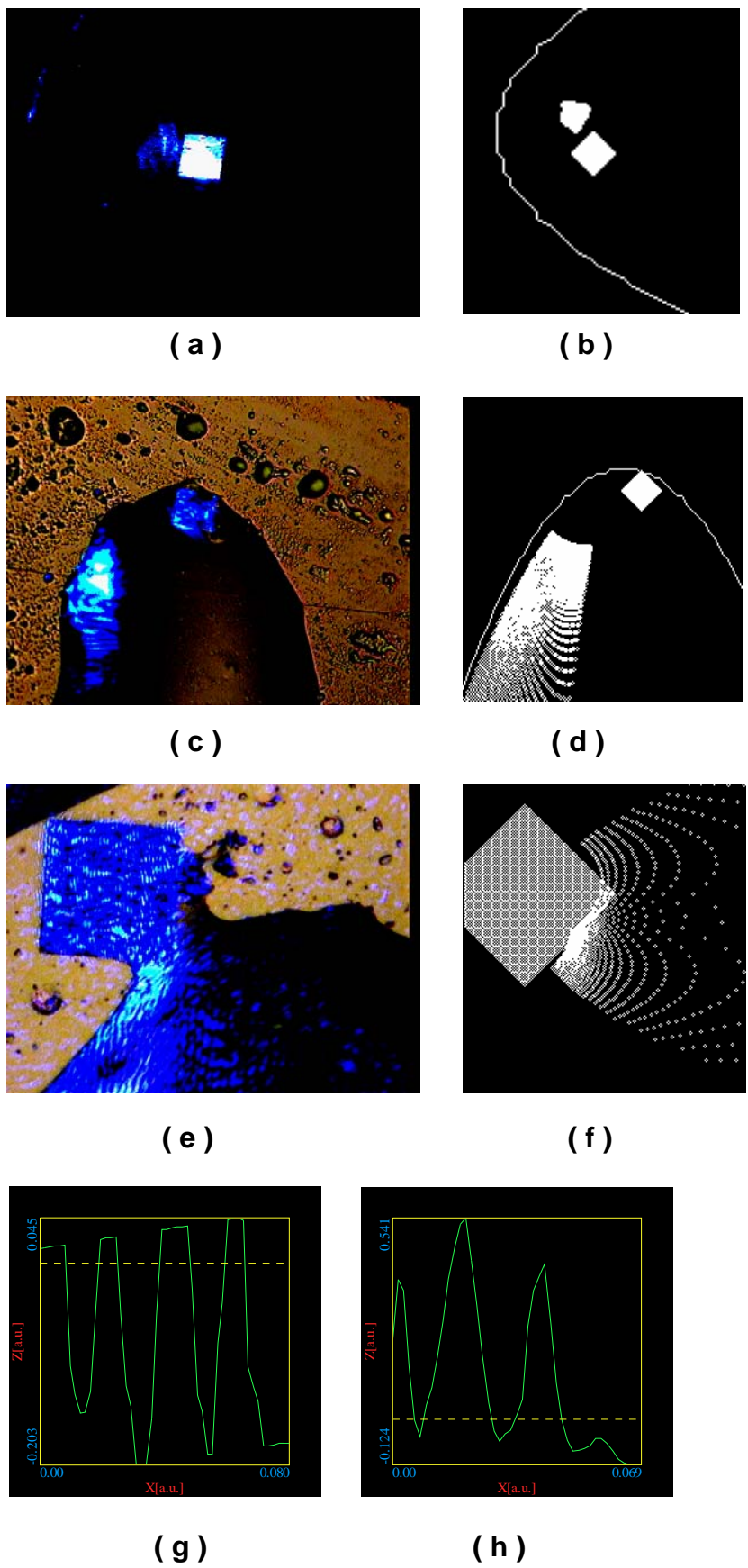

Fig.3

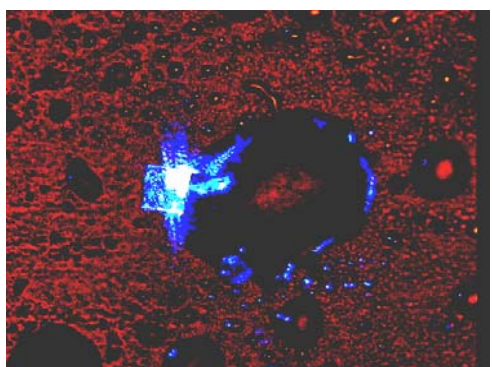
(a)

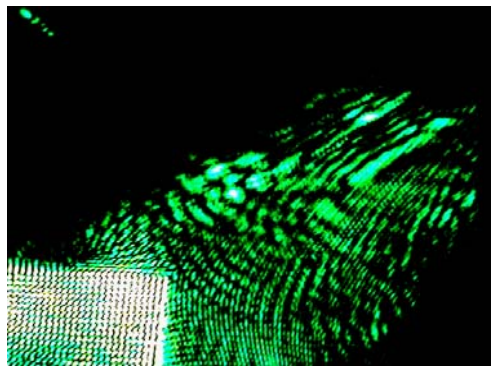
(b)

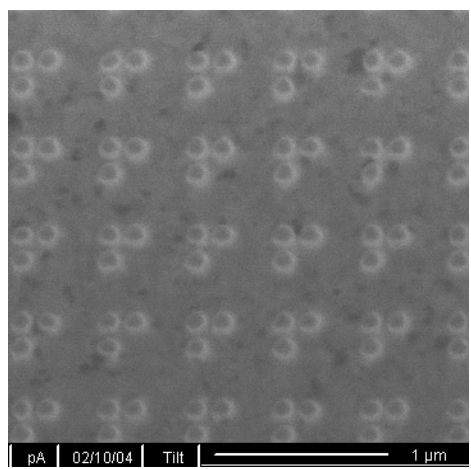
(c)

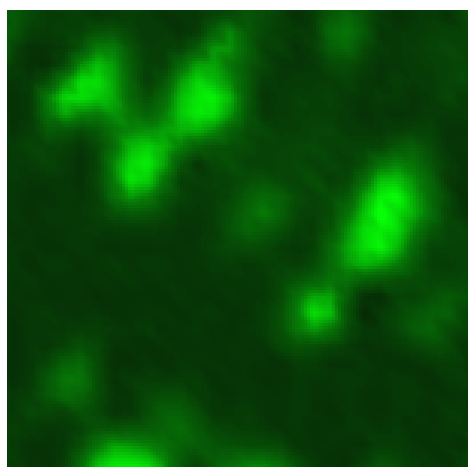
(d)

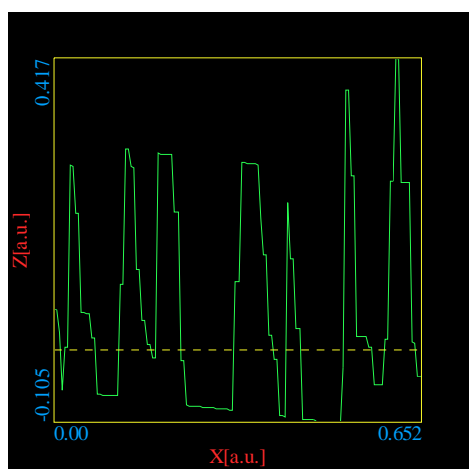
(e)

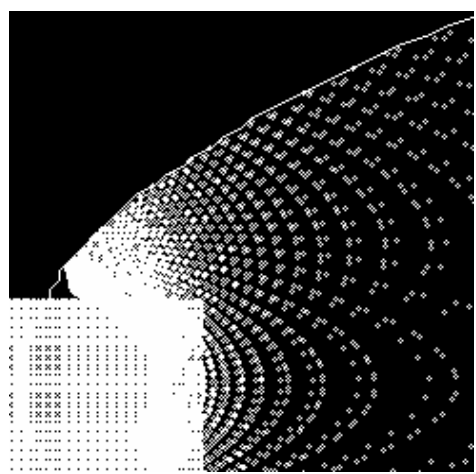
(f)

Fig. 4